\documentclass[prb,reprint]{revtex4-1}
\usepackage{graphicx}
\usepackage{subfigure}
\usepackage{amsmath}

\begin{document}


\title{A simple table-top demonstration of radiation pressure on a macroscopic object}


\author{G. Jesensky}
\affiliation{Dept. of Physics, 901 12th Ave. Seattle University, Seattle, WA 98122}
\author{D. Dams}
\affiliation{Dept. of Physics, 901 12th Ave. Seattle University, Seattle, WA 98122}
\author{O. Khomenko}
\affiliation{Dept. of Physics, 901 12th Ave. Seattle University, Seattle, WA 98122}
\author{W. J. Kim}
\affiliation{Dept. of Physics, 901 12th Ave. Seattle University, Seattle, WA 98122}
\affiliation{Center for Experimental Nuclear Physics and Astrophysics (CENPA), University of Washington, Seattle, WA 98195}


\date{\today}

\begin{abstract}
We report a simple demonstration of radiation pressure on a table-top experiment. Utilizing dynamic force microscopy in ambient environment, the resonant motion of a cm-sized cantilever driven by an amplitude-modulated diode laser is directly observed. Our versatile setup involves a host of exciting techniques that are relevant in precision force measurements and represents an ideal experiment in the undergraduate laboratory. 
\end{abstract}


\maketitle


\section*{Introduction}
Radiation pressure represents one of the fundamental aspects of James Clerk Maxwell's electromagnetic theory. In 1873, on the basis of his theory, Maxwell predicts that ``by means of the concentrated rays from an electric lamp falling on a thin metallic disc, delicately suspended in a vacuum, it is possible to produce an observable mechanical effect.''\cite{max} Thirty years later in 1903, Ernest Fox Nichols and Gordon Ferrie Hull at Dartmouth College conducted torsion balance measurements to demonstrate the effect of radiation pressure on a suspended mirror for the first time; their experiment, though successful, was challenging due to spurious effects caused by the motion of gas molecules and direct heating of radiation itself.\cite{hull} 

More than 100 years have passed since the landmark experiment by Nichols and Hull, and in 2009 the Radiation Pressure Prize Challenge was officially announced, calling for ``a simple, direct, and educational low-cost experimental demonstration of the effect of radiation pressure.''\cite{challenge} Further, the guidelines stipulates that ``the effect should be seen with simple means without the need of sophisticated data analysis (a directly visible effect is preferred).'' 

The winner of the call has not yet been announced as of writing this report, let alone publications reporting an experiment meeting all of the requirements. It is true that radiation pressure is routinely measured and even exploited in other areas of physics such as optomechanics in which direct observation of the ground-state of the mechanical motion of a macroscopic object is one of the primary goals.\cite{opto1,opto2} Yet, all of these experiments require micro- or nano-fabricated resonators in high-vacuum, low-temperature conditions. Additionally, the minuscule size of the resonators employed in these experiments often makes it difficult to distinguish between two competing driving mechanisms: radiation pressure and photo-thermal effects.\cite{therm,munday}   

Although our experiment described in this report is not intended to be a submission in response to the call,\cite{deadline} we believe that our simple, direct approach to measure the effect of radiation pressure satisfies the aforementioned conditions and makes an exciting addition to the undergraduate laboratory. The basic idea behind our experiment is simple: Cut out a rectangular bar out of common metal, find its natural frequency $\nu_0$, and drive it with a laser pointer modulated near $\nu_0$ in ambient environment. When this simple idea is properly executed, one would expect a resonance curve of a damped driven harmonic oscillator peaked around its resonant frequency, with the realization that the driving agent is now the laser pointer exerting a real mechanical force on an everyday object.  

Our experiment is composed of three distinct stages, and we shall discuss them in a chronological order in which they are performed. The three stages are (I) Michelson's interferometry, (II) dynamic force microscopy, and (III) radiation-driven resonance. Each stage provides an experimental platform by which a unique scientific topic of its own can be explored. Much of the experimental work described in this report has been conducted by one undergraduate student in an 8 week-period of summer research in 2015, with some help of two students as they pursues their own independent research projects.   

\section*{Stage I: Michelson interferometer} 
The first stage of our experiment is construction of a motion sensor. For resonance detection in the context of dynamic force microscopy (to be discussed in next section), one could either use a fiber optic interferometer \cite{rugar,kim1,DFM} or a Michelson interferometer (MI). We chose the latter, because we had previously exploited its usefulness for another undergraduate experiment \cite{hankins} and its integration into the undergraduate laboratory is commonplace nowadays with a number of reports describing the MI technique in details.\cite{nachman,caltech}    

Shown in Fig. 1 is a schematic of our MI (top) and the result of its metrology (bottom) for determining the wavelengths of two different laser sources. 
\begin{figure}[htpb]
\centering
\includegraphics[width=0.7\columnwidth,clip]{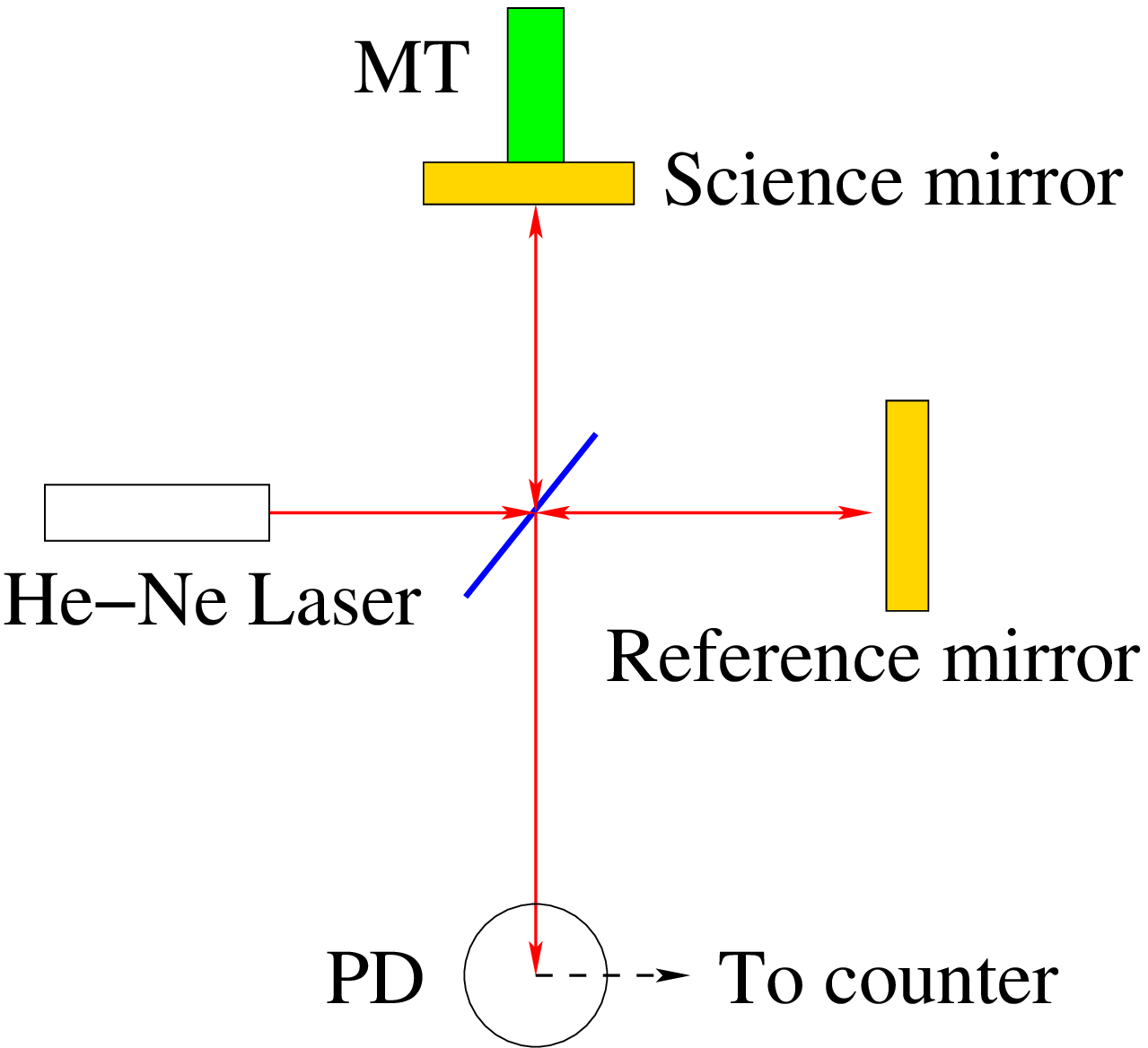}
\includegraphics[width=0.8\columnwidth,clip]{green_red.eps}
\caption{Schematic of our Michelson interferometer (top) and the result of fringe counting of two different laser sources, green and red (bottom): A laser source is split by a beam splitter, which then follows two different paths, one reflecting off the ``science mirror" and the other off the ``reference mirror''. When the science mirror actuated by a motorized transducer (MT) moves a distance $d$ causing the path length difference of $2d$, the intensity of the interference created by the recombined beams changes by one period across a photo-detector (PD). The ``peaks'' are then counted by a counter interfaced with a National Instrument DAQ card (NI-USB-6251). The number of fringe counts ($\Delta N$) is inversely proportional to the change in distance ($\Delta d$) through the relation: $\Delta N/\Delta d=2/\lambda$. The two ``unknown'' laser sources are correctly identified, and their wavelengths are resolved to be $\lambda_{\rm{red}}=635.0\pm 0.1$ nm and $\lambda_{\rm{green}}=542.2\pm 0.2$ nm.}
\label{fig1}
\end{figure}
Briefly, a laser source is split by a beam splitter, reflected by two orthogonal mirrors, and recombined to create constructive and destructive fringe patterns, collected by a photo-detector (PD). The fringe counts $N$ is inversely proportional to the path length difference $2d$, twice the change in distance caused by a motorized transducer (MT) as it moves the ``science'' mirror, with the relationship given by $N=2d/\lambda$. By measuring the slope $\Delta N/\Delta d$, one can determine the ``unknown'' wavelength $\lambda$ of the laser source. The two distinct slopes (i.e., red and green lasers) are easily resolved as an offset in the log-log plot, as demonstrated in the bottom of Fig. 1.

To put our results in perspective, a change of $d$=0.1 mm for the green laser makes about $N=300$ peaks to pass by on the detector. Fig. 2 shows a close-up view of fringe patterns displaying 4 peaks over 10 $\mu$m change in distance. Note that the spacings between two successive data points are somewhat irregular; this is because the separation between two successive points has nearly reached the displacement resolution of the MT, which is about 50 nm.
\begin{figure}[htpb]
\centering
\includegraphics[width=0.9\columnwidth,clip]{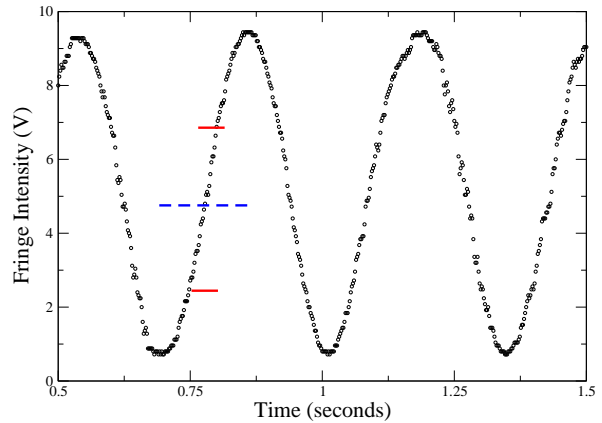}
\caption{Close-up view of fringe patterns detected on the PD: The science mirror moves about 10 $\mu$m over a one-second interval, inducing $N=$4 peaks to be counted at the counter. Our motorized transducer (Thorlabs Z825V) has resolution no better than 50 nm, and this limitation shows up as distortions in the data points.The dotted blue line represents a ``sweet'' spot of our interferometer where the maximum sensitivity is attained, a feature fully exploited in our DFM stage.}
\label{fig2}
\end{figure}

The fringe counting procedure has been fully automated in a MATLAB script using the counter/timer function of the National Instrument DAQ card. The measurements involving the movement of the MT and counting the number of fringes on the PD takes no more than 5 minutes. More importantly, the successful metrology of MI ensures that various optical components are properly aligned and provides important calibrations for dynamic force measurements discussed in our next stage.  

\section*{Stage II: Dynamic force microscopy} 
The science mirror is now replaced by a cantilever with its one end clamped; the cantilever is cut from a brass bar and its surfaces, front and back, are optically polished, stepping down to 0.3 micron grit optical polishing paper. Its rectangular shape has dimension $w=(6.42\pm0.02)$ mm, $l=(11.81\pm0.02)$ mm, $t=(0.85\pm0.02)$ mm, width, length, and thickness, respectively. The cantilever, via its clamping device, is attached to a piezoelectric transducer (PZT), which acts as a driving source.  
 \begin{figure}[htpb]
\centering
\includegraphics[width=0.7\columnwidth,clip]{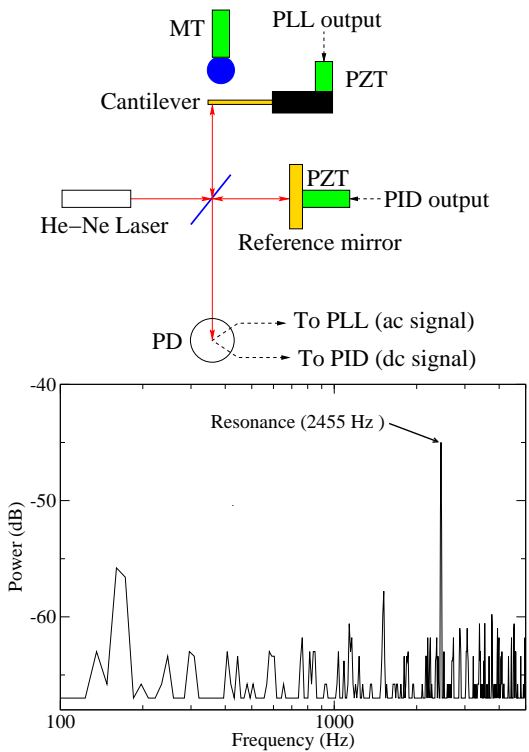}
\includegraphics[width=0.8\columnwidth,clip]{fft.eps}
\caption{Schematic of our dynamic force microscopy setup (top) and detection of a resonant peak on the fast-Fourier spectrum (botton): The science mirror is now replaced by a surface-polished cantilever of which motion is monitored in a phase-locked loop (PLL). The PLL essentially tracks the motion of the cantilever at its resonant frequency $\nu_0$ by constantly adjusting a piezoelectric transducer attached onto the clamping device of the cantilever. Note that the reference mirror is attached with another PZT; a proportion-integration-derivative (PID) controller (Stanford Research: SIM960) actively translates this reference-PZT to optimally position the fringe location at its maximum sensitivity (e.g. half-way between the constructive and destructive peaks). The PID output is based on the error signal between its set point and the PD signal with a time constant of order of a second.}
\label{fig3}
\end{figure}
 
\subsection{Resonance detection}
To detect the resonant motion of the cantilever, a phase-locked loop (PLL) is implemented: A lock-in amplifier (EG\&G 7260 DSP) receives ac-signals containing the resonant motion of the cantilever ($\nu_0$) from the PD and drives the PZT at a new frequency ($\nu_0'$) that gives the maximum response. Unlike a typical lock-in amplifier in which signals are processed at a single fixed frequency, the PLL dynamically adjusts its frequency to provide the maximum response. This is useful because the natural frequency of our cantilever constantly changes due to temperature drifts, air current, vibration and etc.

The bottom of Fig. 3 shows a resonant peak detected around $\nu_0=2455$ Hz on a fast Fourier-transform (FFT) spectrum. While the resonance captured on the FFT spectrum is a snap-shot, the PLL constantly tracks and updates the resonant motion of the cantilever. The resonant frequency, once captured by the PLL, remains locked within 0.1 Hz of the peak for as long as several hours.  

Another feature of our DFM setup is a proportion-integration-derivative (PID) loop that involves the reference mirror to which another PZT is attached. The purpose of implementing the PID loop is to ensure that the interferometric ``path'' stay at its most sensitive position, as indicated by the blue line in Fig. 2, where the slope of the intensity change is the maximum. The positions capped by the two red lines indicate a typical amplitude range of our PID, $\Delta V_{\rm{PID}}=5 $ V, with its time constant $\tau_{\rm{PID}}\approx 1$ s. In contrast, the amplitude of the PLL output is much smaller $\Delta V_{\rm{PLL}}\le 10 $ mV at a significantly faster rate $1/\nu_0$=500 $\mu$s. Essentially, the reference mirror is moving slowly over a wider range of the interference path, whereas the science mirror (e.g., cantilever) is moving rapidly over a narrower range of the path. 

In principle, one could proceed with radiation pressure measurements at this point by replacing the mechanical-driving source (PZT) with an intensity-modulated light source without performing dynamic force measurements. In doing so, however, two important issues might be overlooked: First, because the cantilever is clamped to another mechanically rigid structure, which is also driven by the PZT, other resonances that are unrelated to the actual motion of the cantilever exist. Only by shifting the resonant frequency of the cantilever with a known external force could one be assured that the resonance detected in Fig. 3 is a true natural frequency of the cantilever itself; second, one of the critical parameters characterizing the cantilever is its effective mass $m_{\rm{eff}}$. Performing dynamic force measurements provides direct means to quantify this parameter through electrostatic calibrations.      

\subsection{Dynamic force detection}
To enable dynamic force measurements, we establish an electrostatic force by applying a voltage $V$ across the grounded cantilever and a brass sphere of $R=7.93\pm0.01$ mm, placed on the back side of the cantilever (see Fig. 3). The gap distance $d$ between the cantilever and the sphere can be adjusted by the MT attached to the sphere. 

When a cantilever of natural frequency $\nu_0$ experiences an external force $F(d)$, the theory of DFM\cite{kim2} predicts that the resonant frequency of the cantilever shifts by an amount: $\Delta\nu^2\equiv\nu^2-\nu_0^2$, where $\nu$ is the shifted frequency due to the external force. For an attractive force, the amount of shift is negative ($\Delta\nu^2<0$); and for a repulsive force, it is positive ($\Delta\nu^2>0$). If the force is known, the total amount of frequency shift can be directly calculated by taking the gradient of the force:
\begin{equation}
\Delta\nu^2=-\frac{1}{4\pi^2 m_{\rm{eff}}}\frac{\partial F(d)}{\partial d}.
\end{equation}  
In the case of our sphere-plane geometry, the electrostatic force is given by $F_{\rm{elect}}=-\pi\epsilon_0 RV^2/d$, where $\epsilon_0$ is the vacuum permittivity constant. Thus, the electrostatic contribution to the total frequency shift is then
\begin{equation}
\Delta\nu_{\rm{elect}}^2=-\frac{\epsilon_0R}{4\pi m_{\rm{eff}}d^2}V^2.
\end{equation}  
At a given distance $d$, the amount of frequency shift is a parabolic function of applied voltages: That is, $\Delta\nu_{\rm{elect}}^2\propto V^2$. The negative sign indicates the attractive nature of the electrostatic force. 

Demonstrated in top of Fig. 4 are the observed curves of the frequency shifts as a function of $V$ at two different gap locations: $d_{\rm{large}}$(a) and $d_{\rm{short}}$(b), where the measurement at $d_{\rm{large}}$ is taken at 3 times further out from a contact point than at $d_{\rm{short}}$. At the shorter distance, the amount of frequency shifts has drastically increased ($\Delta\nu^2$ going from -1000 to -5500 Hz$^2$), even with a significant reduction in applied voltages (from $\pm$25 V to $\pm$1 V). This reflects the power law nature ($1/d^2$) of the strength of the electrostatic gradient, as predicted by Eq. (2).  
\begin{figure}[httpb]
\centering
\includegraphics[width=0.9\columnwidth,clip]{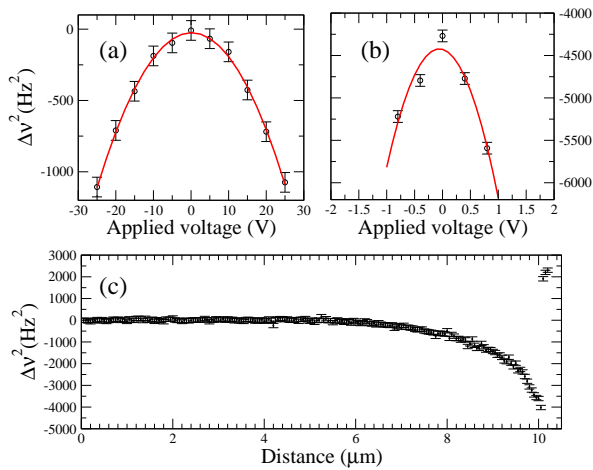}
\caption{Results of dynamic force microscopy: The frequency-shift is plotted as a function of applied voltages at two different locations: $d_{\rm{large}}$(a) and $d_{\rm{short}}$(b). Based on parabolic fits and using Eq. (2), the absolute separating distances are estimated to be $d_{\rm{large}}=(262\pm1$) $\mu$m and  $d_{\rm{short}}=(8\pm1$) $\mu$m and the effective mass $m_{\rm{eff}}=(5.0\pm0.5)\times10^{-8}$ kg. The bottom plot is a graph of force versus distance when no voltage is applied across the cantilever and the sphere. Even with electrostatic voltage applied, significant frequency shifts are evident reaching $\Delta \nu^2\approx-4000$ Hz$^2$ right before contact.}
\label{fig4}
\end{figure}

A couple of remarks are in order with regard to our electrostatic measurements: First, the minima of our parabola measurements do not necessarily occur at $V=0$, as predicted by Eq. (2); instead, an offset voltage $V_0$ must be applied to achieve the minimized electrostatic interaction. That is, $V^2\rightarrow (V-V_0)^2$, where $V=V_0$ attains the minimum of the electrostatic interaction. In literature, this offset voltage is known as a volta or contact potential and ubiquitously exists between metals due to imperfect surface conditions.\cite{kim2} Based on our parabolic measurements, we obtain a contact potential $V_0=100\pm2$ mV, consistent with the observed offsets reported elsewhere. 

Second, to obtain the absolute separating distance $d$, we impose $d\equiv d_r-d_0$ by introducing another fit parameter $d_0$ into Eq. (2). This parameter represents the asymptotic limit of the $1/d^2$-power law, also known as the point of contact. Note that $d_r$ is the relative distance recorded by our translation stage MT. For example, the relative distances recorded for the large and short measurements were $d_r^{\rm{large}}$=0.9360 mm and $d_r^{\rm{short}}$=0.6824 mm. The second-order polynomial coefficient in Eq. (2) $a\equiv\Delta\nu^2_{\rm{elect}}/(V-V_0)^2$ is thus expressed as
\begin{equation}
a=-\frac{\beta}{(d_r-d_0)^2},
\end{equation} 
where $\beta\equiv\epsilon_0R/4\pi m_{\rm{eff}}$. Based on the two polynomial coefficients obtained from our parabola measurements, we obtain the point of contact $d_0=(0.674\pm0.001) $mm and $\beta=(1.121\pm0.003)\times10^{-7}$ $\rm{Hz^2m^2/V}^2$, yielding the actual values for the large and short distances to be $d_{\rm{large}}=(262\pm1$) $\mu$m and  $d_{\rm{short}}=(8\pm1$) $\mu$m, and the effective mass $m_{\rm{eff}}=(5.0\pm0.5)\times10^{-8}$ kg. 

Third, the sole contribution of electrostatic interaction at its minimum, even with an offset voltage, should yield no frequency shift (e.g., $\Delta \nu_{\rm{elect}}^2$=0), as predicted by Eq. (2). Yet our data show an significant amount of frequency shifts, with $\Delta\nu^2$ at the electrostatic minimum approaching -4200 Hz$^2$ at the short distance measurement. This suggests an additional contribution from a force of non-electrostatic origin, such as the Casimir force, owing to the short-ranged nature of electromagnetic interactions between the employed objects. To see this clearly, we plot the progression of the frequency-shift as the sphere is brought into contact with the cantilever at zero electric potential $V=0$ (Fig. 4c). Despite the possible existence of residual electrostatic interactions, a significant amount of the frequency shift is observed over 2-3 $\mu$m distances right before making a hard contact in the repulsive regime.\cite{note1}

In summary, the advantage of performing DFM in the context of our radiation pressure measurement is two fold. One is the unambiguous identification of the resonant frequency of a resonator by shifting its frequency in dynamic force measurements; the other is to characterize the key parameter like the effective mass of the resonator through electrostatic calibrations. Together, it is possible to determine the degree to which the resonator responds to the actual force of radiation pressure. 
  
\section*{Stage III: Radiation-driven resonance}
The radiation pressure experiment is now performed by driving the cantilever with an intensity-modulated (or ``chopped'') laser. The modulated light source strikes the back surface of the cantilever and drives it near resonance, as shown in the top of Fig. 5. 
\begin{figure}[htpb]
\centering
\includegraphics[width=0.7\columnwidth,clip]{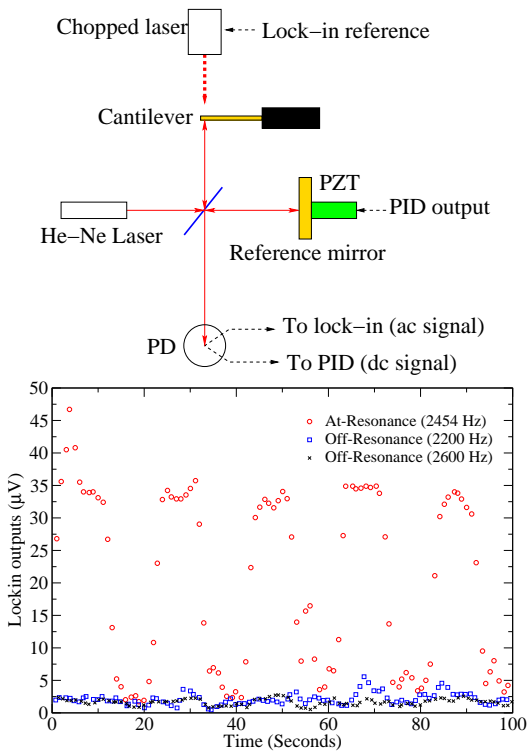}
\includegraphics[width=0.8\columnwidth,clip]{at_resonance2.eps}
\caption{Schematic of the radiation-driven resonance setup (top) and the lock-in outputs corresponding to response amplitudes at three different frequencies (bottom). The PZT that used to drive the cantilever the PLL is now removed, and a second light source chopped near the resonance is introduced; the chopping frequency provides the reference signal for the standard lock-in amplification. A clear contrast is seen between the lock-in response at resonance and those at off-resonance. The light beam is physically blocked at every 20 seconds to rule out any electrically-driven resonance.}
\label{fig5}
\end{figure}
The chopping frequency is set manually by the reference frequency of the lock-in amplifier. The PID loop is maintained to maximize the interferometer's sensitivity. 

In the bottom of Fig. 5, we present the lock-in response when the modulated light is striking the cantilever at resonance ($\nu_0=2455$ Hz) and at off-resonances ($\nu_{\pm}=2455\pm200$ Hz). To decouple from possible spurious pickups from electronics, the light path is physically blocked every 20 seconds. A huge response at resonance is evident demonstrating the radiation-driven resonance. At off-resonances, the responses between ``block'' and ``unblock'' are indistinguishable. Next, we repeat the procedure by recording the lock-responses in terms of the difference between when the light is blocked and unblocked (or the lock-in difference) at several different frequencies near the resonance. The result is plotted in Fig. 6.
\begin{figure}[htpb]
\centering
\includegraphics[width=0.9\columnwidth,clip]{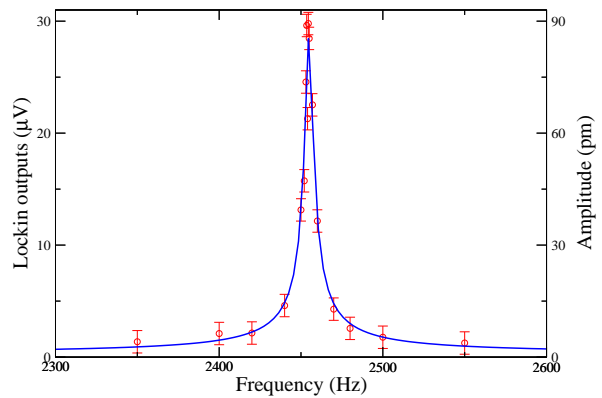}
\caption{Radiation-driven resonance: Data points (red,circle) are fit to the Lorentzian function of a form: $A(\nu)=a+b/\sqrt{(\nu-\nu_0)^2+\gamma^2}$. Based on the fit, the resonance is found at $\nu_0=(2454.9\pm0.3)$ Hz, with the quality factor $Q=\nu_0/2\gamma=505\pm81$. Other fit parameters are: $a=0.13\pm$0.50 $\mu$V and $b=35\pm5$ $\mu$V/Hz. The lock-in responses are calibrated with known displacements caused by the driving PZT employed in Fig. 3. The converted scales are shown on the vertical axis on the right-hand side.}
\label{fig6}
\end{figure}
The amplitude versus frequency of oscillation, as measured by the lock-in difference at different frequencies, is fit to the Lorentzian response function for a damped, driven harmonic oscillator,\cite{taylor} with the resonant frequency and the quality factor determined to be $\nu_0=(2454.9\pm0.3)$ Hz and $Q=505\pm81$.

The amount of force exerted by the radiation pressure can be calculated from an equation: $F_{\rm{opt}}=(2R+A)P_0/c$, where $c$ is the speed of light, $P_0$ the output power of the modulated light source (30 mW), and $R=0.4$ and $A=0.6$, respectively, the reflectivity and absorptivity coefficients measured for the polished surfaces of our cantilever. Together, the force $F_{\rm{opt}}$ is estimated to be 140 pN. Based on the effective spring constant 11.9 N/m of the cantilever obtained from the DFM measurements (e.g., $k_{\rm{eff}}=m_{\rm{eff}}\nu_0^2$), the displacement caused by the light source at $\nu<<\nu_0$ (or off-resonance) is $x=F_{\rm{opt}}/k_{\rm{eff}}=1.2\times10^{-11}$ m or 12 pm. This falls near our detection limit, which is about 9 pm (see Fig. 6). At resonance, the radiation-driven response is enhanced by an order of magnitude and peaks around 90 pm, which is readily resolved in our data.  

Note that our signal-to-noise ratio is greater than 10. Even with optical power of our light source reduced by a factor of six (or $P_0^{\rm{min}}$=5 mW), the resonant peak would have been still resolved at around 15 pm, well above the noise floor of 9 pm. Indeed, a simple commercial laser pointer would be sufficient for driving our cantilever.  

One of the primary competing mechanisms that mimic the effect of radiation pressure is the photo-thermal effect.\cite{therm,munday} To see if this effect is responsible for the resonance observed in our experiment, we have measured the thermal time constant $\tau$ for our cantilever by heating it up and measuring the time it takes to cool down to an equilibrium temperature. The result is presented in Fig. 8. The measured thermal constant $\tau\approx$ 40 sec is too slow a recovery time compared to the much faster, mechanical response rate to the modulation: $1/\nu_0 \approx500$ $\mu$sec.
\begin{figure}[htpb]
\centering
\includegraphics[width=0.9\columnwidth,clip]{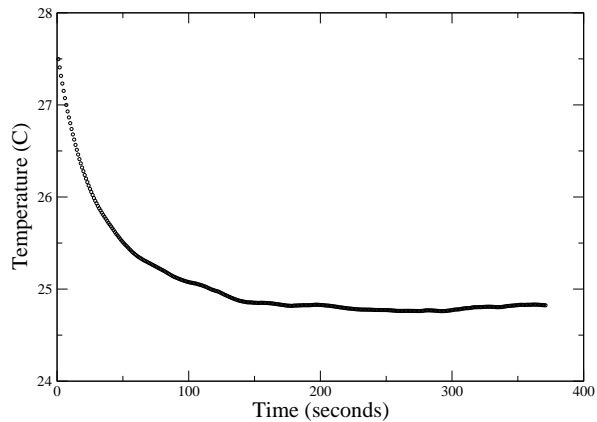}
\caption{Measurement of the thermal time constant of the cantilever: The cantilever is briefly heated to a few degrees above the room temperature and relaxed to cool down to an equilibrium. A thermistor, attached on the ``back'' side of the cantilever, monitors the temperature change on the surface of the cantilever. The data are fit to a functional form $T=T_0+Be^{-t/\tau}$, with $\tau$ (thermal time constant) estimated to be $40.2\pm0.3$ seconds. Other fit parameters are: $T_0=24.8\pm0.5$ $^{\circ}$C and $B=2.56\pm0.01$ C$^{\circ}$.}
\label{fig7}
\end{figure}
Unlike many of the radiation-pressure experiments involving micro- or nano-fabricated resonators in which the photo-thermal effect plays a significant role due to the short thermal constant on the order of msec to $\mu$sec,\cite{therm,munday} the use of a macroscopic (cm-sized) object as employed in this report makes the distinction between the two mechanisms much easier. 

\section*{Conclusion}
We have reported a simple, table-top experiment to measure the effect of radiation pressure in ambient environment. A cm-sized cantilever cut from a piece of ordinary bulk metal can be driven to resonance by a light source, with an optical power as low as that available in a laser pointer. Our experiment makes an exciting addition to the undergraduate laboratory as it involves a variety of useful experimental techniques ranging from Michelson's interferometry, to dynamic force microscopy, and to resonance detection using electronic feedback systems, such as PLL and PID loops. The techniques surrounding dynamic force microscopy can be readily expanded to explore short-ranged, electromagnetic interactions between a pair of metallic objects, also an exciting topic of contemporary physics immediately accessible to the undergraduate laboratory.    

\begin{acknowledgments}
The authors acknowledge financial support from the Clare Boothe Luce Undergraduate Research Scholars Program, Murdock Trust through the Murdock College Science Research program, and the National Science foundation through RUI/PHY-1307150. They also thank Chris Varney for preparing the surface-polished cantilever used in this report. 
\end{acknowledgments}


\begin{thebibliography}{99}
\bibitem{note1} Although the Casimir force is an interesting subject of its own, analyzing it based on our short-range dynamic force measurements is beyond the scope of the present report.
\end{thebibliography}

\end{document}